\documentclass{elsart}
\usepackage{graphicx}
\journal{PHENO 2001}

\begin{document}
\begin{frontmatter}
\hfill$\vcenter{\hbox{\bf ULB-TH/01-13}}$

\title{Prompt Atmospheric Neutrinos: 
\newline Phenomenology and Implications\thanksref{pheno}}
\vspace{.1in}
\author{{\sc C.G.S. Costa} and {\sc C. Salles}}
\thanks[pheno]{Talk presented at the {\it PHENO 2001 Symposium}, 
May 7-9, 2001, \newline University of Wisconsin - Madison.
http://pheno.physics.wisc.edu/pheno01/}
%\thanks[ead]{e-mail: cecosta@ulb.ac.be}
\address{Service de Physique Th\'eorique, CP 225, 
Universit\'e  Libre de Bruxelles, 
\newline Boulevard du Triomphe, 1050 Brussels, Belgium}
\date{May 25, 2001}
\vspace{.5in}
\begin{abstract}
We present an overview of the phenomenology of prompt neutrinos 
produced in the atmosphere by the decay of charmed particles. 
In face of the great activity involving operational and proposed 
large scale neutrino telescopes, we discuss the implications of 
different scenarios in calculating the prompt flux, upon the 
detectable signals in these experiments.
\end{abstract}
\end{frontmatter}
\newpage
\section{Introduction}
One could imagine to build a special kind of experimental 
facility to study high-energy and very-high-energy neutrinos. 
While on the wishful thinking, one could gather all the best 
conditions. Desirable characteristics include:
\begin{itemize}
\item to provide acceleration of parent particles up to 
the highest energies;
\item to be coupled to the largest particle beam dump;
\item to be shielded by a set of the largest and most 
dense absorbers available; 
\item to be monitored by an ensemble of several detectors, 
each operated with a different detection technique, 
and complementary acceptances;  
and 
\item to produce and detect neutrinos covering a broad 
range of energies, perhaps from $10^{6}$eV to $10^{18}$eV.
\end{itemize}

Does it sounds like a too difficult enterprise?
Well, we get very close to it, if we regard all cosmic-ray 
phenomena and corresponding experiments in a complementary way. 

Think of several cosmic particle acceleration mechanisms, 
provided by sources such as supernova explosions, neutron stars in 
binary systems, active galactic nuclei, gamma-ray bursts, pulsars, 
rapidly spinning magnetars and yet some exotic objects 
(topological defects, decaying massive relics). 
Choose as target our planet, where gravitationally 
trapped gaseous elements serve as a beam dump, and the 
solid core material, from one surface to the other, 
provide at the same time shielding and support for 
different purpose detectors, scattered around at different 
heights and depths. 
A block diagram of this ``ultimate'' high-energy facility 
looks like:

\vspace*{1cm}
\[ \hspace*{-1.4cm}
\fbox{$\begin{array}{c}
Cosmic \\ Accelerator 
\end{array}$ }
\rightarrow 
\fbox{$\begin{array}{c}
Atmospheric \\ Beam \;Dump 
\end{array}$ }
\rightarrow 
\fbox{$\begin{array}{c}
Planetary \\ Rock \;Shield 
\end{array}$ }
\rightarrow 
\fbox{$\begin{array}{c}
Complementary \\ Detectors
\end{array}$ }
\]

\begin{figure} %[t]
\begin{center}
\vspace*{13cm}
\hbox{ %\hspace{5cm} 
        \includegraphics{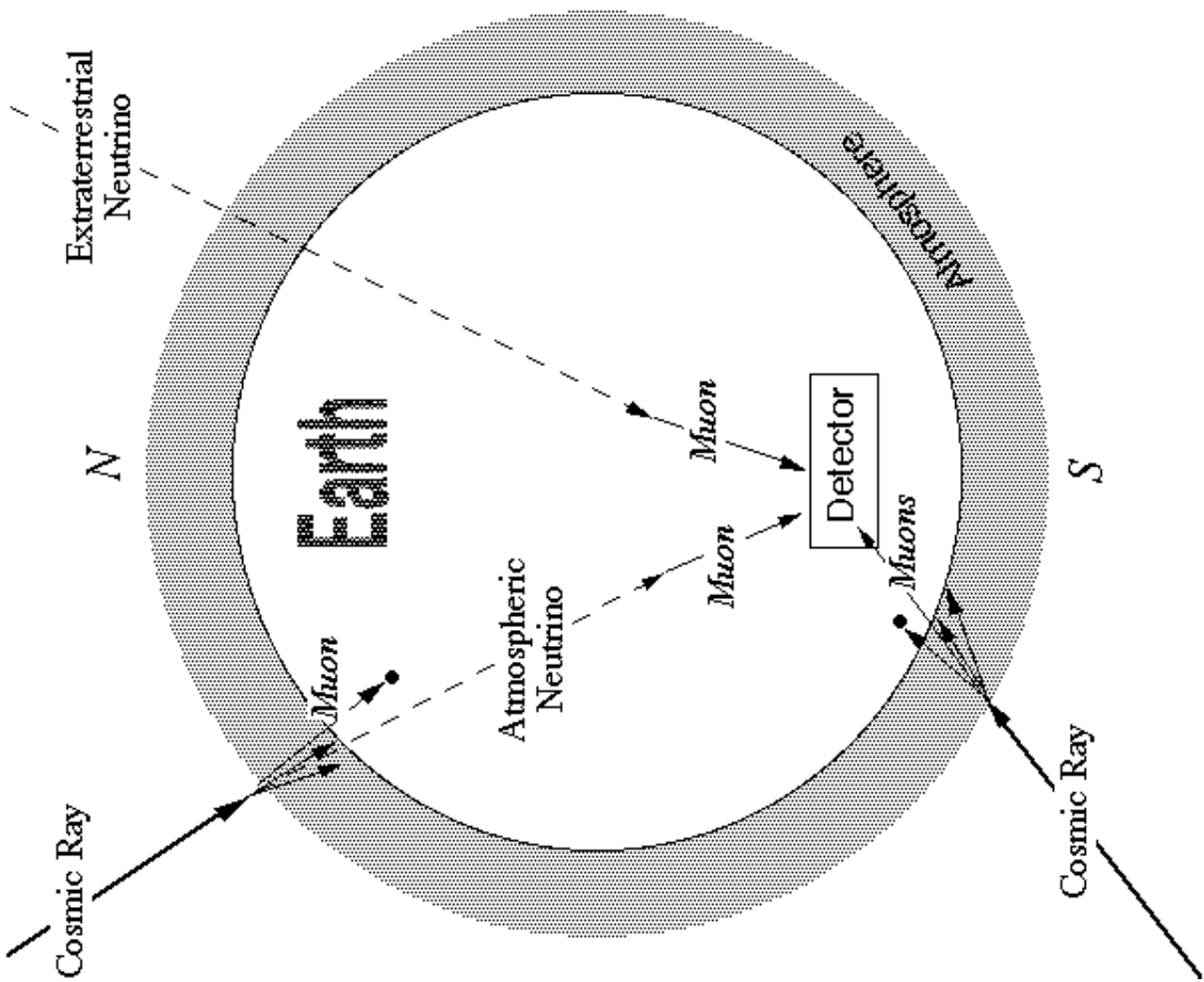}}
\end{center}
\caption[]{Sketch of the ``ultimate'' experimental neutrino 
facility (the Cosmos included!).}
\label{fig:device}
\end{figure}

\newpage
A more realistic sketch for this omnipresent experimental neutrino 
facility is presented in Figure~\ref{fig:device}. One should 
consider the whole cosmos as part of the figure. One may also 
include several detectors scattered around, with different and 
complementary purposes, for example: 
optical and gamma-ray space telescopes; 
primary cosmic-ray satellite and balloon borne calorimeters; 
mountain-altitude cosmic-ray family emulsion chamber detectors; 
cherenkov, fluorescence and particle extensive air shower detectors; 
muon magnetic spectrometers; and yet 
large underground, underwater and under-ice observatories. 
%
% FIGURE 1
%%%%%%%%%%%%%%%%%%%%%%%%%%%%%%%%%%%%%%%%%%%%% FACILITY

High-energy extraterrestrial neutrinos (above tens of TeV) 
may be directly produced at the same (galactic or extra-galactic) 
sources of high-energy cosmic-ray particles and gamma-rays, including 
those of the highest energies. 
Low energy neutrinos (below tens of MeV) are produced directly 
in nuclear reactions at the core of the Sun. 

Intermediate energy neutrinos are produced locally, at the 
planetary scale. A cosmic accelerated primary particle, a nucleon 
for example, impinging upon nuclei of air in the atmosphere, 
will generate a shower of secondary particles, which may propagate 
to the ground or even below. The main contributions to air showers 
come from the prodution of a) neutral pions, giving rise to 
electromagnetic cascades (a sequence of photons and 
electron-positron pairs); and of b) charged mesons 
(mainly pions and kaons), giving rise to hadronic cascades.
Charged-meson decays will give rise to neutrinos and muons; 
these muons on their turn may eventually decay into neutrinos, 
all contributing to what is called the {\it conventional 
atmospheric neutrino flux}\cite{conventional}. 

The semileptonic decay of heavy quark particles, produced in the 
hadronic cascades initiated by cosmic-ray interactions, 
also gives rise to atmospheric neutrinos, a particular set known as 
the {\it prompt atmospheric neutrino flux}. 

These prompt neutrinos are the subject of the present talk. 
Among the motivations for their study, we highlight the following:
\begin{itemize}
\item High Energy Neutrino Telescopes - 
\\ Though neutrinos are the most abundant cosmic rays at sea level, 
their measurement requires large underground detectors,  
in order to compensate for the small neutrino-nucleon cross section. 
Nowadays operational neutrino telescopes, like 
AMANDA\cite{amanda} and BAIKAL\cite{baikal}, are taking and 
analizing data. They are being optimized to increase signal, 
data statistics and energy coverage, and to reduce the background, 
mainly caused by muons and by electronics limitations. 
Other similar and much larger instruments are 
being contemplated (Antares, Nestor, Nemo, IceCube)\cite{telescopes}. 
As will be shown later, there is the possibility that prompt 
neutrinos are observable in such neutrino detectors\cite{windowpaper}. 
\item Background to Cosmic Neutrinos - 
\\ The measurement of galactic and extra-galactic neutrinos, 
which is the main purpose of neutrino telescopes, is of course 
limited by discrimination against atmospheric neutrinos, 
the prompt component included. 
\item Bounds to Charm Production Cross Section - 
\\ The actual measurement of prompt neutrinos, or the failure to 
do so, will impose constrains to the charm production cross section 
at high energy\cite{isrpaper}, in a similar way as it has been done 
through the investigation of muon poor horizontal air 
showers\cite{halzen:94}.
\item Proton Small-x Gluon PDF -
\\ It has been shown that the spectral index of the prompt atmospheric 
neutrino flux may depend linearly on the slope of the gluon distribution 
function at very small $x$, not reachable at colliders\cite{ggv}. 
\item Background to High Energy Neutrino Oscillations - 
\\ Prompt decay of the charm-strange meson $D_{s}$ provide the unique 
{\it direct} source of high-energy tau-neutrinos in the atmosphere, which may 
compete with those coming from flavor oscillation of atmospheric 
or extraterrestrial muon-neutrinos\cite{windowpaper}. 
\end{itemize}

Why charm turns out to be of such particular interest 
to the study of high energy neutrinos? 
\newline
The answer comes easily from the analysis of the 
{\it critical energy} of different neutrino-parent particles 
produced in cosmic-ray interactions in the atmosphere. 
The critical energy delimits the competition between decay and 
interaction lengths, that is to say, 
above this energy the parent particle is likely to interact or 
be slowed down, rather than to decay into a neutrino. 
%Table~\ref{table:pdg} 
Table~1 compares the critical energies 
calculated for muons, pions and kaons (parents of the conventional 
atmospheric neutrino flux) and for relevant charm baryons and 
mesons (parents of the prompt flux). While the contributions of 
the conventional component to the neutrino flux are constrained below 1~TeV, 
the charm prompt decays - although less copious - represent the 
{\it only} atmospheric contribution up to around 100,000~TeV. 

In what follows we present an outline of the calculation of the 
prompt neutrino component in the atmosphere, comparing different 
phenomenological approaches found in the literature. 
Next we discuss the implications of the different prompt flux 
scenarios upon the detectable signals in large scale neutrino 
telescopes.

\vspace*{1cm}
% TABLE 1
%**********************************PDG 
%\begin{table}[hb]   
\begin{table}[hb]  
\caption[]{\label{table:pdg}  
Critical energies $\varepsilon_{critic}$ of selected particles.}     
\smallskip 
\centering 
\begin{tabular}{lcrc}      
\hline   
Particle &Elementary &$mc^2$ &$\varepsilon_{critic}$\\  
&contents &(MeV)&(GeV)\\  
\hline 
$\mu^{+},\mu^{-}$ &lepton              &106 &1.0\\   
$\pi^{+},\pi^{-}$ &$u\bar{d},\bar{u}d$ &140 &115\\ 
$K^{+},K^{-}$ &$u\bar{s},\bar{u}s$     &494 &855\\ 
$\Lambda^{o}$  &$uds$  &1116 &$9.0 \times 10^4$\\ 
\hline     
$D^{+},D^{-}$ &$c\bar{d},\bar{c}d$ &1870 &$3.8 \times 10^7$\\  
$D^{o},\bar{D}^{o}$ &$c\bar{u},\bar{c}u$ &1865 &$9.6 \times 10^7$\\  
$D_{s}^{+},D_{s}^{-}$ &$c\bar{s},\bar{c}s$ &1969 &$8.5 \times 10^7$\\   
$\Lambda_{c}^{+}$  &$udc$  &2285 &$2.4 \times 10^8$\\  
\hline  
\end{tabular}   
\end{table}   

\newpage
\section{Calculation}
The particle and energy flow along the development of an air 
shower, at different atmospheric depths $x$ (given in units 
g/cm$^{2}$), relevant for the production of prompt leptons, 
can be summarized as:
\[
\begin{array}{cc}
\fbox{$\begin{array}{c}
Primary \\ Cosmic \;Ray
\end{array}$}
&
\Phi_{N} (E_{N},x=0) \: = \: N_{o} \:
E_{N}^{-(\gamma+1)}
\\
\downarrow\downarrow\downarrow & \\
\fbox{$\begin{array}{c}
Nucleonic \\ Flux
\end{array}$}
&
\Phi_{N}(E_{N},x)\: = \: N_{o} \:
E_{N}^{-(\gamma+1)} \:
e^{-x/\Lambda_{N}} 
\\
\downarrow\downarrow & \\
\fbox{$\begin{array}{c}
Charm \;Production \\ and \;Decay
\end{array}$}
&
\:\: d\Phi_{i}/dx \:(E_{i},x)\: = \:
-\frac{1}{\lambda_{i}}\Phi_{i}  -\frac{1}{d_{i}}\Phi_{i}
+ K_{i} e^{-x/\Lambda_{N}} 
\\
\downarrow & \\
\fbox{$\begin{array}{c}
Prompt \;Lepton \\ Flux
\end{array}$}
&
\Phi_{l}(E_{l},x_{l})
         \:=\: \int dx 
         \:\: \int dE_{i} 
         \:\: \frac{df^{l}}{dE_{l}} \:\: 
         B_{i} \: \frac{1}{d_{i}} \: 
         \Phi_{i}(E_{i},x)
\end{array}
\]

The index $N$ stands for the nucleons (protons, neutrons) arriving 
at the top of the atmosphere ($x=0$), with differential energy 
spectrum $\Phi_{N}$ described by a power law. They cascade down, 
subjected to an attenuation length $\Lambda_{N}$, to a generic 
level $x$ where they may produce a charmed particle of type-$i$ 
($i = D^{\pm}, D^{o}, \bar{D}^{o},  D_{s}^{\pm}, \Lambda_{c}^{+}$). 
The term $K_{i}$ is related to the charm production 
spectrum-weighted moment $Z_{Ni}$, while $\lambda_{i}$ and $d_{i}$ 
are respectively the charm interaction and decay lengths. 
Finally, upon decay of the charmed particle to produce the 
lepton $l$, with branching ratio $B_{i}$ and decay spectrum 
$df^{l}/dE_{l}$, we have calculated the differential flux of 
prompt leptons $\Phi_{l}(E_{l},x_{l})$, with energy $E_{l}$ at 
atmospheric depth $x_{l}$.

Detailed calculation of the prompt lepton component is presented 
in Ref.\cite{cookbook}. In order to have a taste of what this 
result may look like, and what it signifies, we write down 
an approximate solution, valid for energies below the charm critical 
energy and for depths of the the order of the whole atmosphere: 
\[
\begin{array}{lcccc}
\Phi_{l} (E) \approx &Z_{i\rightarrow l}(E) & \times\; Z_{N\rightarrow i}(E) &
\times\; \left(\Lambda_{N} /\lambda_{N}\right) 
& \times\; \Phi_{N} (E,0) \\
\rm{Prompt \;Lepton} &  
\leftarrow \rm{Decay} &  
\leftarrow \rm{Produce} &  
\leftarrow \rm{Attenuate} &  
\leftarrow \rm{Primary \;Flux} 
\end{array}
\]

Reading from right to left, this expression tells us what we have 
seen in the previous diagram: the primary flux is attenuated, 
then produces charm particles which decay into prompt leptons 
(the decay spectrum-weighted moment $Z_{il}$ is related to the 
decay spectrum $df^{l}/dE_{l}$).

Does it seems that simple? Well, we must first take a look at the 
many ingredients used in the calculation, considering that there 
are many possible phenomenological assumptions to be adopted for each. 

\vspace{-0.5cm}
\section{Phenomenology}
\vspace{-0.5cm}
A preliminary list of parameters involved in the cascade routines 
leading to the calculation of the prompt flux comprises:
\begin{itemize}
\item Primary Spectrum Amplitude $N_{o}$ and Slope $\gamma$
\item Nucleonic Inelastic Cross Section $\sigma_{in}^{\! N-air} (E)$
\item Nucleonic Interaction Length $\lambda_{N} (E)$
\item Nucleonic Spectrum-weighted Moment $Z_{NN}(\gamma,E)$
\item Nucleonic Attenuation Length $\Lambda_{N} (E)$
\item Charm Inelastic Cross Section $\sigma_{in}^{\! i-air} (E)$
\item Charm Interaction Length $\lambda_{i} (E)$
\item Charm Production Spectrum-weighted Moment $Z_{Ni}(\gamma,E)$
\item Three-body Semileptonic Decay Spectrum $\frac{df^{l}}{dE_{l}}$
\item Charm Decay Spectrum-weighted Moment $Z_{il}(\gamma)$
\item Charm Decay Length $d_{i}$
\end{itemize}

It is up to the user to choose a desired brand 
for each ingredient. 
As an example, consider some charm production models 
found in the literature: 
\begin{description}
\item[Quark Gluon String Model (QGSM) - ]
A semi-empirical model of charm production based on the 
non-perturbative QCD calculation by Kaidalov and 
Piskunova\cite{kaidalov:86}, normalized to accelerator data, 
and applied to the prompt muon calculation by Volkova 
{\it et al.}\cite{qgsm}. 
\item[Recombination Quark Parton Model (RQPM) - ]
A phenomenological non-per\-turb\-a\-tive approach, 
taking into account the contribution of the intrinsic charm to
the production process, in which a $c\bar{c}$ pair is coupled to
more than one constituent of the projectile hadron, as described
by Bugaev {\it et al.}\cite{rqpm}. 
\item[Perturbative QCD (pQCD) - ]
There are several calculations based on perturbative QCD, 
for example Thunman {\it et al.}\cite{tig:96} evaluate explicitly 
the charm production up to leading order (LO) in the coupling 
constant, using the Monte Carlo program PYTHIA, and include 
the next-to-leading order (NLO) distribution effects as an 
overall factor. More recently Gelmini {\it et al.}\cite{ggv:00} 
updated the calculation to include the full contribution of NLO 
predictions to the lepton fluxes. Pasquali {\it et al.}\cite{prs:99} 
offer an alternative implementation of the perturbative QCD approach.  
\item[Limiting Conditions - ]
E. Zas {\it et al.}\cite{halzen:93} calculate extreme cases of 
charm production, at both low and high production rate limits.  
Higher rates are achieved assuming a charm production cross 
section which is 10\% of the total inelastic cross section 
(called Model-A), extrapolated with a $\log^2 (s)$ energy dependence. 
Lower rates are obtained using pQCD at NLO, with structure 
functions given by Kwiecinski-Martin-Roberts-Stirling and adopting 
relatively hard parton distribution functions (called Model-E). 
\end{description}

\begin{figure} 
\begin{center}
\vspace*{20cm} 
\hbox{ 
        \includegraphics{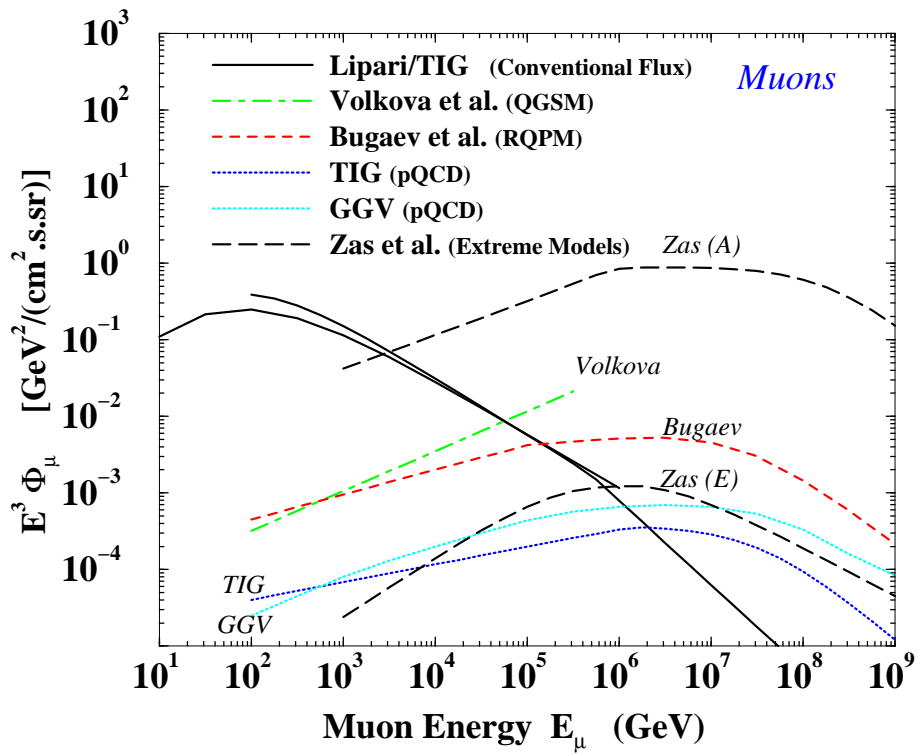}} 
\end{center}
\caption[]{Comparison of calculated differential vertical 
atmospheric muon fluxes at sea level, as reported by several 
authors (see text).}
\label{fig:reported}   
\end{figure}

Prompt muon fluxes calculated assuming these charm production models 
are compared in Figure~\ref{fig:reported}. (The fluxes are 
multiplied by $E^{3}$ in order to flatten the curves.) 
Two calculations of the conventional atmospheric muon flux are also 
included, and we note that
the cross-over of the conventional and prompt components may occur 
at any energy between $2 \times 10^3$ GeV and $2 \times 10^6$ GeV.
The prompt flux intensity, itself, spans up to four orders of
magnitude! However, we call attention to the fact that the 
comparison is somewhat unfair, since not only different charm 
production models are being compared: calculations of different 
authors involve different assumptions for the other cascading 
ingredients, so that we are seeing a combination of effects.  

% FIGURE 2
%**********************************REPORTED

\begin{figure} 
\begin{center}
\vspace*{8cm} 
\hbox{ 
        \includegraphics{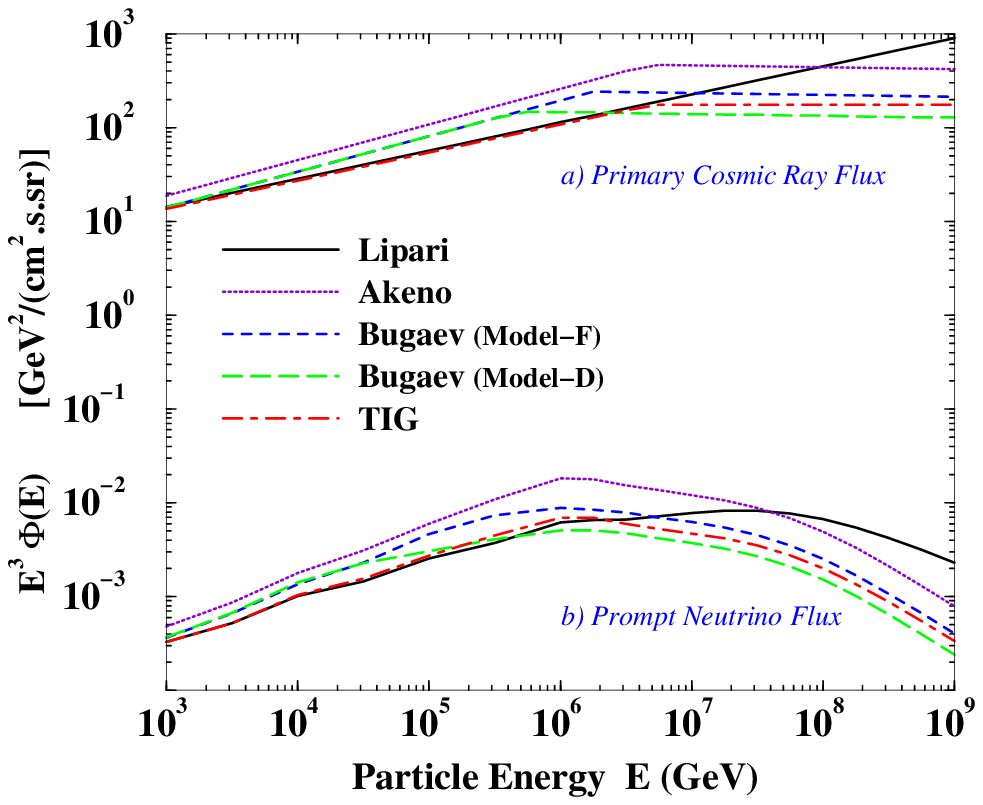}} 
\end{center}
\caption[]{a) Comparison of primary cosmic-ray energy spectra, as  
given by different parametrizations ; 
b)comparison of corresponding prompt neutrino fluxes,   
assuming all the ingredients fixed but the primary spectrum.}
\label{fig:primes}   
\end{figure}

\begin{figure} 
\begin{center}
\vspace*{9cm} 
\hbox{ 
        \includegraphics{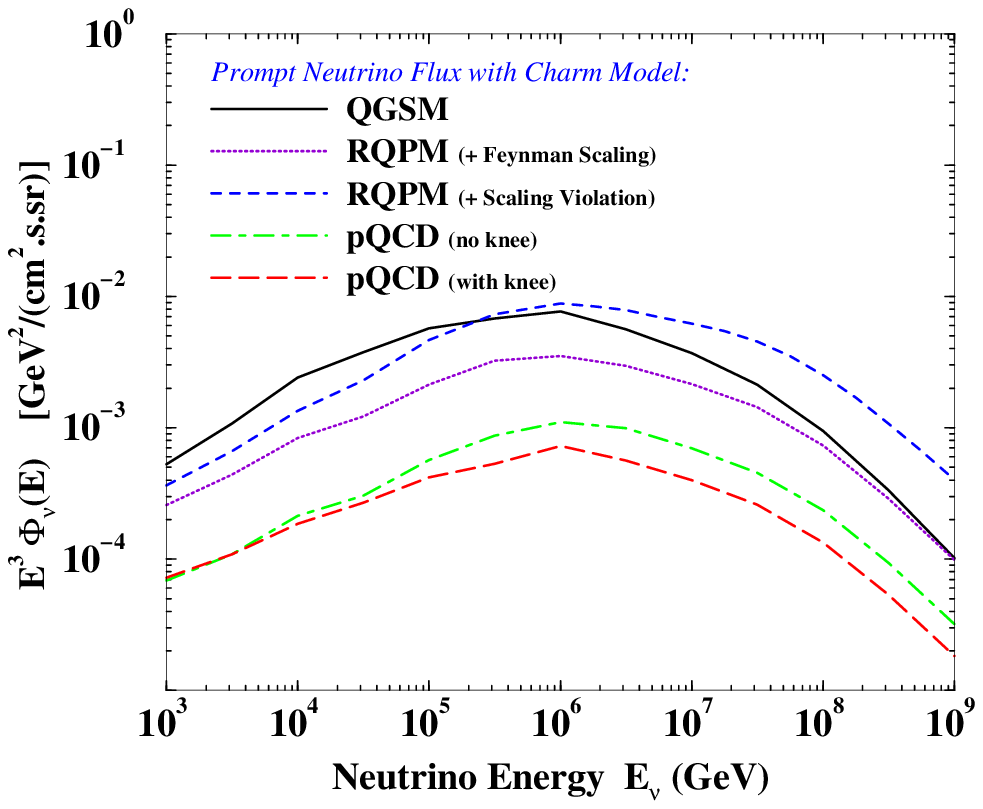}} 
\end{center}
\caption[]{Comparison of prompt neutrino fluxes  
for different charm Z-Moment models, 
assuming all the other ingredients fixed.}
\label{fig:charm}   
\end{figure}

Ref.\cite{cookbook} discusses the influence of each separate 
ingredient upon the final prompt flux, adopting diverse 
parametrizations available in the literature. 
For example, the solely effect of the primary cosmic-ray spectrum 
is displayed in Figure~\ref{fig:primes}. Curves in a) show the 
spectrum at the top of the atmosphere, according to 5 models. 
Curves in b) are the corresponding prompt muon-neutrino vertical 
flux at sea level.
Same comparison can be made, fixing all ingredients but the charm 
production spectrum-weighted moment. The resulting muon-neutrino 
fluxes for 5 different parametrizations of charm model are shown 
in Figure~\ref{fig:charm}.  
 
% FIGURE 3
%**********************************RESULTS:Primes

% FIGURE 4
%**********************************RESULTS:Charm

\begin{figure} 
\begin{center}
\vspace*{8cm} 
\hbox{ 
        \includegraphics{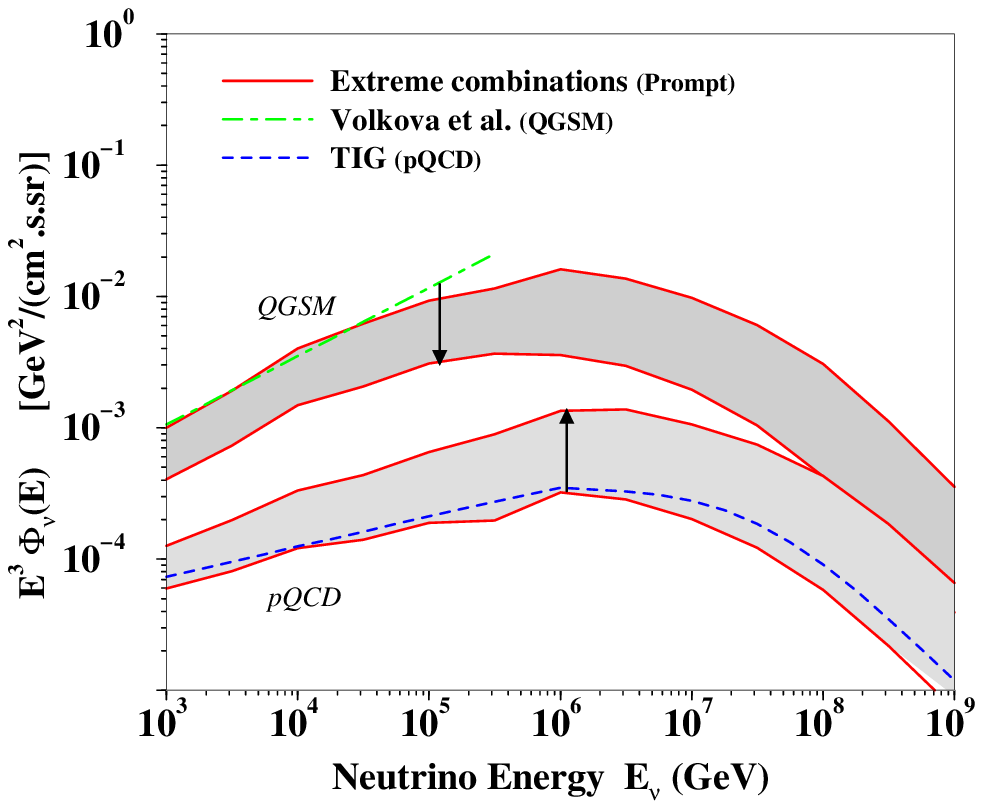}} 
\end{center}
\caption[]{Prompt neutrino flux calculated with QGSM (upper band) 
and with pQCD (lower band). 
The bands represent the range between extreme ingredient combinations, 
for each charm model. Prompt flux from Volkova {\it et al.} 
(dotted line) and from Thunman {\it et al.} (dot-dashed line) 
also shown for illustration.}
\label{fig:shift1}   
\end{figure}

\begin{figure} 
\begin{center}
\vspace*{9cm} 
\hbox{ 
        \includegraphics{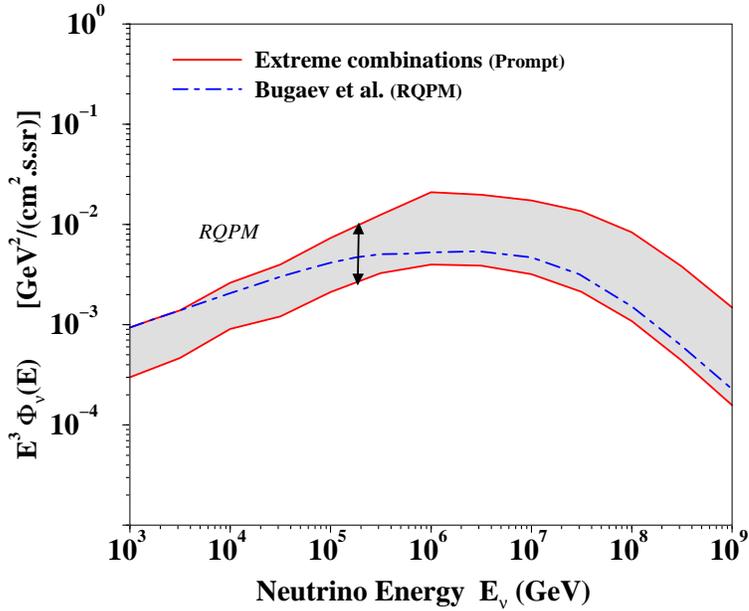}} 
\end{center}
\caption[]{Prompt neutrino flux calculated with RQPM. 
The band represents the range between extreme ingredient combinations. 
Prompt flux from Bugaev {\it et al.} (dot-dashed line) 
also shown for illustration.}
\label{fig:shift2}   
\end{figure}

We can think now of mixing different parametrizations of all 
ingredients listed previously. It is possible to choose extreme 
combinations, that is to say, those leading to lower and higher 
prompt lepton fluxes, for a given charm model. 
This procedure can alter the flux which has been reported in the 
literature: Figure~\ref{fig:shift1} shows how the calculation of 
Volkova {\it et al.} can be shifted down and how the calculation 
of Thunman {\it et al.} can be shifted up, by adopting different 
ingredients, other then the charm model. With the same procedure, 
Figure~\ref{fig:shift2} shows how the calculation 
of Bugaev {\it et al.} wiggles! 

% FIGURE 5
%**********************************RESULTS: Shift QGSM+pQCD

% FIGURE 6
%%**********************************RESULTS: Shift RQPM

\newpage
\section{Implications}
\vspace{-0.5cm}
We have seen how the prompt flux calculation is subjected to large 
uncertainties, resulting from the imprecise knowledge of both 
atmospheric particle showering parameters and the modeling of 
the production of charm. 
We observe that the calculated neutrino fluxes may be shifted 
by up to one order of magnitude for a given charm model. 
When comparison is made among alternative charm production models, 
the spread reaches two orders of magnitude. 
We will therefore define an allowed range between the maximum (MAX) 
and minimum (MIN) prompt neutrino fluxes. The MAX combination of 
ingredients adopts the QGSM below the knee and the RQPM above, 
in addition to other extreme parameters. 
The MIN combination adopts the pQCD charm model, with ingredients 
that pulls the flux down (complete description of choices is listed 
in Ref.\cite{cookbook}). 

It is not difficult to extend the previous calculation to obtain 
the flux of prompt tau-neutrinos, the main contribution coming 
from decay of $D_{s}$-mesons into a $\tau\nu$-pair, and subsequent 
in-chain tau decays producing more neutrinos\cite{pr:99}. 
The above mentioned conclusions on the uncertainties of the 
calculation still hold for the tau component. 

Applying the prescription of Gandhi {\it et al.}\cite{gandhi}, 
we can use the calculated fluxes (of prompt $\nu_{\mu}$ and 
$\nu_{\tau}$) to estimate the rates  
of neutrino-induced muons and taus, per year per effective detection  
area (in $2\pi$ sr), in operating and proposed neutrino telescopes. 
Table~\ref{table:rates} summarizes the event rates for 
BAIKAL NT-200\cite{baikal}, which have an effective area 
of $2\times 10^{3}$~m$^{2}$, 
for AMANDA-II\cite{amanda},
with effective area of $3\times10^{4}$~m$^{2}$, and for two 
different thresholds in the proposed km$^{3}$ experiment 
ICECUBE\cite{ice3}. The estimates for the conventional 
atmospheric neutrino component (CONV) are shown for comparison. 
Experimental signatures may come to the rescue on disentangling 
the two components. For example, the $\approx$ 1,500 prompt 
$\nu_{\mu}$'s above 1 TeV in ICECUBE data may be extracted 
using their flat zenith angular distribution 
(while the conventional flux has a strong angular dependence). 
Prompt $\nu_{\tau}$'s would produce characteristic showers 
in the detector, enabling discrimination of the $\approx$~200~(20) 
neutrino induced taus above 1~(10)~TeV. 

\begin{figure}
\begin{center}
\vspace*{10cm} 
\hbox{ 
        \includegraphics{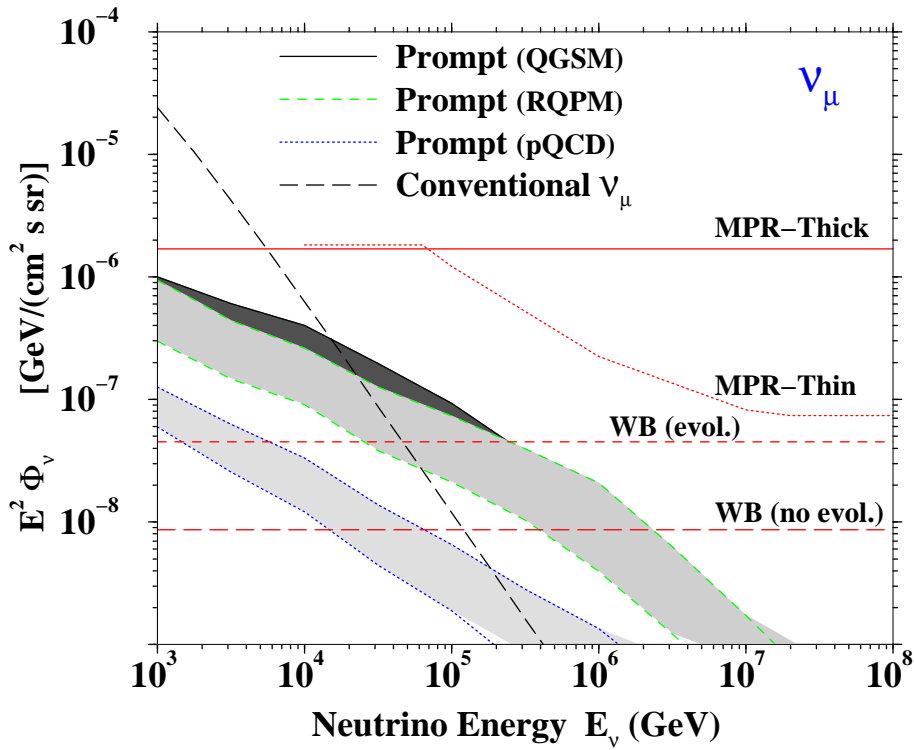}} 
\end{center}
\caption[]{Comparison of several contributions to
the high-energy $\nu_{\mu}$ flux. The bands represent the
allowed range between  maximum and minimum atmospheric 
prompt neutrino fluxes for different charm production models. 
Thick dashed line is the conventional atmospheric flux.
Solid, dotted, dashed and long-dashed lines correspond to upper
bounds imposed to diffuse extragalactic neutrino flux by the
observation of high-energy cosmic-ray and gamma-ray spectra,
according to different scenarios explained in the text.
Fluxes are multiplied by $E^{2}$.}
\label{fig:window1}
\end{figure}

The next question is how the conventional and prompt atmospheric 
fluxes compare to expected rates of extragalactic 
neutrinos\cite{windowpaper}. Adopting for comparison the upper 
bound to the diffuse flux of high-energy neutrinos from cosmic 
accelerators, such as active galactic nuclei and gamma-ray bursts, 
we summarize the situation for muon-neutrinos in Figure~\ref{fig:window1}. 
The bands are the allowed region for prompt neutrinos as in 
Figures~\ref{fig:shift1} and~\ref{fig:shift2}. 
Also shown is the conventional atmospheric flux. 
Concerning extraterrestrial neutrinos, the two upper curves,  
labeled MPR (solid and dotted lines), represent the bounds imposed 
to neutrino fluxes from sources that are optically thick or thin to 
the emission of
neutrons\cite{mpr}, respectively. 
The two straight lines labeled WB (dashed and long--dashed), 
represent bounds on sources which produce the highest energy 
cosmic rays\cite{wb}. The higher flux allows for cosmological 
evolution of the sources. 
We observe, from Figure~\ref{fig:window1}, that the prompt neutrino 
flux exceeds the conventional at some energy above 20~TeV. 
If the MPR bounds are to hold, the prompt component is merely 
a background to be discounted. However, according to the WB-scenario, 
the prompt component may exceed the extragalactic diffuse flux up 
to 300~TeV, if the sources are to evolve, and up to 2~PeV if not. 

\begin{figure}
\begin{center}
\vspace*{9cm} 
\hbox{ 
        \includegraphics{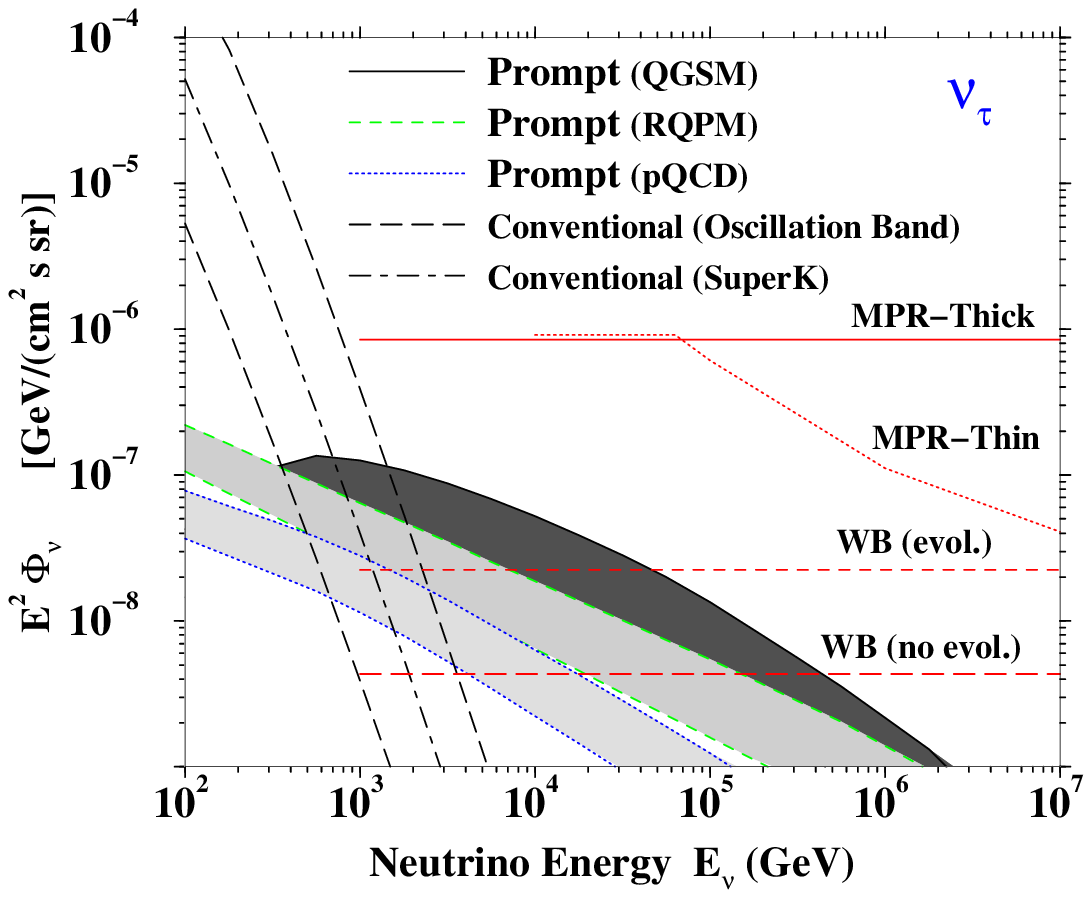}} 
\end{center}
\caption[]{Comparison of several possible contributions
to the high-energy $\nu_{\tau}$ flux. Again, the bands represent the
allowed range between maximum and minimum atmospheric 
prompt neutrino fluxes for different charm models.
Thick dashed lines are the result of maximum mixing flavor
oscillation of the conventional atmospheric $\nu_{\mu}$ flux, for
$\Delta m^{2}$ around the Super-Kamiokande value 
$3.2 \times 10^{-3}$~eV$^{2}$(thick dot-dashed
line). 
Solid, dotted, dashed and long-dashed lines correspond to the 
upper bounds in Figure~7, subjected to vacuum flavor oscillation, 
averaged in transit to Earth.}
\label{fig:window2}
\end{figure}

Allowing for flavor oscillation of atmospheric and extragalactic 
muon-neutrinos  into tau-neutrinos, Figure~\ref{fig:window1} can be 
transformed into Figure~\ref{fig:window2}, remarking that prompt 
$\nu_{\tau}$'s are directly produced from $D_{s}$ decays and 
are independent of oscillation mechanisms. 
Once again we observe the possibility that there is an 
observational window opened for the prompt component, 
this time for tau-neutrinos, extending from 2~TeV up to 40~TeV 
(or up to 500~TeV, depending on the scenario). 
% 
% TABLE 2
%**********************************Table: Event Rates. 
\begin{table}[h] 
\caption[]{\label{table:rates} Upward-going muon and tau 
event rates per year arising from charged current interactions of 
atmospheric neutrinos in ice or water, for different neutrino 
telescope's effective areas and thresholds.  
Prompt flux calculation based on maximum (MAX) and minimum (MIN) 
range limits described in the text, compared to conventional  
atmospheric flux (CONV).} 
\smallskip 
\centering 
\begin{tabular}{l|lll|ll} 
\hline\hline 
           & \multicolumn{3}{c}{$\mu^{+}+\mu^{-}$ } 
           & \multicolumn{2}{c}{$\tau^{+}+\tau^{-}$}\\ 
Experiment &CONV     
& Prompt & Prompt & Prompt & Prompt \\ 
           && MAX & MIN & MAX & MIN \\ \hline 
BAIKAL NT-200    
&22    &3    &0   &0   &0 \\ 
AMANDA-II       
&330   &44   &2   &7   &0 \\ 
ICECUBE ($>1$ TeV)  
&11000 &1470 &53  &216 &11 \\ 
ICECUBE ($>10$ TeV)   
&170   &157  &3   &22  &1  \\ 
\hline\hline 
\end{tabular} 
\end{table}

%%%%%%%%%%%%%%%%%%%%%%%%%%%%%%%%%%%%%%%%%%%%%WINDOW 1 
%
%\newpage
% FIGURE 7
%
%%%%%%%%%%%%%%%%%%%%%%%%%%%%%%%%%%%%%%%%%%%%%WINDOW 2  
%\newpage
% FIGURE 8
%

%\newpage
\vspace{-1cm}
\section{Conclusion}  
The calculation of the prompt lepton flux in the atmosphere is 
subjected to large uncertainties. Partially because of our 
imprecise knowledge of certain atmospheric particle showering 
parameters. Further uncertainty is associated with the 
extrapolation to high energy of a variety of models describing 
the accelerator data on charm production. Care must be taken when 
comparing results from different calculations, for the combination 
of ingredients may shift the fluxes up or down. 

Once the uncertainties are quantified, in terms of an allowed band 
for the prompt atmospheric neutrino flux, we come to the result 
that prompt neutrinos may be observable in a kilometer-scale 
neutrino telescope. Actually, it may be possible that a small 
component of prompt neutrinos are already present in the observed 
samples of the ongoing experiment AMANDA, given their current limit 
on the atmospheric neutrino flux\cite{amanda}. 
Of course, the existence of a direct observational window for 
prompt neutrinos depends on the actual bounds to extragalactic 
neutrino fluxes and, in the case of tau-neutrinos, on the actual 
values of the flavor oscillation parameters, which are currently 
subjects of great interest and of intense research. 

The actual measurement of prompt neutrinos, and even the failure to 
do so, will certainly impose constrains to the charm production cross section, 
at energy domains not yet accessible to particle colliders\cite{isrpaper}. 

%And if everything else fails, one can still complain with the owner of the 
%Great Omnipresent Device. 

\section*{Acknowledgements}
%\acknowledgements
The authors would like to thank Jean-Marie Fr\`ere (ULB)
and Daniel Bertrand (IIHE) for support and encouragement. 
CGSC would like to thank Francis Halzen for the kind hospitality 
during the ``PHENO 2001 Symposium'', in Madison, Wisconsin. 
This work was partially supported by the
I.I.S.N. (Belgium) and by The Communaut\'e Fran\c{c}aise de
Belgique - Direction de la  Recherche Scientifique, programme
ARC.

%\end{references}

\end{document}